\documentstyle[12pt]{article}
\def\ie{{\em i.e.}}
\def\eg{{\em e.g.}}

\def\bsg{B(b\to s\gamma)}
\def\beq{\begin{equation}}
\def\eeq{\end{equation}}

\catcode`\@=11 
\def\coeff#1#2{{\textstyle{#1\over #2}}}

\def\lsim{\mathrel{\mathpalette\@versim<}}
\def\gsim{\mathrel{\mathpalette\@versim>}}
\def\@versim#1#2{\vcenter{\offinterlineskip
    \ialign{$\m@th#1\hfil##\hfil$\crcr#2\crcr\sim\crcr } }}
\def\etal{{\em et. al.}}
\def\JL{J. L. Lopez}
\def\DVN{D. V. Nanopoulos}

\def\r#1{$\bf#1$}
\def\rb#1{$\bf\overline{#1}$}

\def\t1{{\tilde 1}}

\def\GeV{\,{\rm GeV}}
\def\TeV{\,{\rm TeV}}

\def\to{\rightarrow}

\def\NPB#1#2#3{Nucl. Phys. B {\bf#1} (19#2) #3}
\def\PLB#1#2#3{Phys. Lett. B {\bf#1} (19#2) #3}

\def\PRD#1#2#3{Phys. Rev. D {\bf#1} (19#2) #3}
\def\PRL#1#2#3{Phys. Rev. Lett. {\bf#1} (19#2) #3}
\def\PRT#1#2#3{Phys. Rep. {\bf#1} (19#2) #3}

\def\TAMU#1{Texas A \& M University preprint CTP-TAMU-#1}

\begin{document}
\begin{flushright}
\baselineskip=12pt
{CTP-TAMU-40/93}\\
{ACT-15/93}\\
\end{flushright}

\begin{center}
\vglue 0.3cm
{\Large\bf The Strongest Experimental Constraints on\\}
\vspace{0.2cm}
{\Large\bf SU(5) x U(1) Supergravity Models\\}
\vglue 0.5cm
{JORGE L. LOPEZ$^{(a),(b)}$, D. V. NANOPOULOS$^{(a),(b),(c)}$,
GYE T. PARK$^{(a),(b)}$,\\}
{and A. ZICHICHI$^{(d)}$\\}
\vglue 0.4cm
{\em $^{(a)}$Center for Theoretical Physics, Department of Physics, Texas A\&M
University\\}
{\em College Station, TX 77843--4242, USA\\}
{\em $^{(b)}$Astroparticle Physics Group, Houston Advanced Research Center
(HARC)\\}
{\em The Woodlands, TX 77381, USA\\}
{\em $^{(c)}$CERN, Theory Division, 1211 Geneva 23, Switzerland\\}
{\em $^{(d)}$CERN, 1211 Geneva 23, Switzerland\\}
\baselineskip=12pt

\vglue 0.4cm
{\tenrm ABSTRACT}
\end{center}
\vglue -0.2cm
{\rightskip=3pc
 \leftskip=3pc
\noindent
We consider a class of well motivated string-inspired flipped $SU(5)$
supergravity models which include four supersymmetry breaking scenarios:
no-scale, strict no-scale, dilaton, and special dilaton, such that only three
parameters are needed to describe all new phenomena $(m_t,\tan\beta,m_{\tilde
g})$. We show that the LEP precise measurements of the electroweak parameters
in the form of the $\epsilon_1$ variable, and the CLEOII allowed range for
$\bsg$ are at present the most important experimental constraints on this class
of models. For $m_t\gsim155\,(165)\GeV$, the $\epsilon_1$ constraint
(at 90(95)\%CL) requires the presence of light charginos
($m_{\chi^\pm_1}\lsim50-100\GeV$ depending on $m_t$). Since all sparticle
masses are proportional to $m_{\tilde g}$, $m_{\chi^\pm_1}\lsim100\GeV$
implies: $m_{\chi^0_1}\lsim55\GeV$, $m_{\chi^0_2}\lsim100\GeV$, $m_{\tilde
g}\lsim360\GeV$, $m_{\tilde q}\lsim350\,(365)\GeV$, $m_{\tilde
e_R}\lsim80\,(125)\GeV$, $m_{\tilde e_L}\lsim120\,(155)\GeV$, and
$m_{\tilde\nu}\lsim100\,(140)\GeV$ in the no-scale (dilaton) flipped $SU(5)$
supergravity model. The $\bsg$ constraint excludes a significant fraction of
the otherwise allowed region in the $(m_{\chi^\pm_1},\tan\beta)$ plane
(irrespective of the magnitude of the chargino mass), while future experimental
improvements will result in decisive tests of these models. In light of the
$\epsilon_1$ constraint, we conclude that the outlook for chargino and
selectron detection at LEPII and at HERA is quite favorable in this class of
models.
}
\vspace{1cm}
\begin{flushleft}
\baselineskip=12pt
{CTP-TAMU-40/93}\\
{ACT-15/93}\\
July 1993
\end{flushleft}
\vfill\eject
\setcounter{page}{1}
\pagestyle{plain}

\baselineskip=14pt

\section{Introduction}
Models of low-energy supersymmetry which have their genesis in physics at
very high energies (\ie, supergravity or superstrings) embody the principal
motivation for supersymmetry -- the solution of the gauge hierarchy problem.
These models can also be quite predictive since their parameters are
constrained by the larger symmetries usually found at very high mass scales.
In contrast, models of low-energy supersymmetry with no such fundamental basis
(such as the minimal supersymmetric standard model (MSSM)) should properly
be acknowledged as generic parametrizations of all possible supersymmetric
beyond-the-standard-model possibilities, and as such not as real ``theories".
Specifically, supergravity models with radiative electroweak symmetry breaking
need only five parameters to describe all new phenomena: the top-quark mass
($m_t$), the ratio of Higgs vacuum expectation values ($\tan\beta$), and
three universal soft-supersymmetry-breaking parameters ($m_{1/2},m_0,A$).
The MSSM on the other hand requires over twenty parameters to describe the
same phenomena, and has led to the erroneous impression that supersymmetric
models necessarily introduce numerous parameters which hamper the analysis
of processes of interest. In order to ameliorate this situation, it is
routinely assumed that these parameters are related among themselves in some
arbitrary way. These ad-hoc simplifications can be misleading (\eg, implying
``general" conclusions based on special cases) or simply insufficient
(\eg, as in the analysis of branching fractions involving particles from all
sectors of the model).

Perhaps one of the most interesting aspects of supergravity models is the
radiative electroweak symmetry breaking mechanism \cite{EWx,LN}, by which the
electroweak symmetry is broken by the Higgs mechanism driven dynamically by
radiative corrections. This mechanism involves several otherwise unrelated
physical inputs, such as the top-quark mass, the breaking of supersymmetry, the
physics at the high-energy scale, and the running of the parameters from high
to low energies. Needless to say, electroweak symmetry breaking has no
explanation in the MSSM (\ie, the negative Higgs mass squared is put in by
hand). In practice, this constraint is used to determine the magnitude of the
Higgs mixing parameter $\mu$ \cite{aspects}, which is of electroweak size or
larger, and implies: (i) a generic correlation among the lighter neutralino and
chargino masses ($2m_{\chi^0_1}\sim m_{\chi^0_2}\sim m_{\chi^\pm_1}$) which has
been observed in a variety of models \cite{ANabc,LNZI,LNZII}, and (ii) the
connection between the supersymmetric sector and the Higgs sector such that, as
the supersymmetry breaking parameters are increased, the Higgs sector
asymptotes quickly to a standard-model-like situation with one light Higgs
scalar \cite{LNPWZh}.

Experimental predictions in this class of models for processes at present and
near future collider facilities can be obtained and have been determined for
the various supergravity models we consider below (LEPI \cite{LNP+LNPZ,LNPWZh},
Tevatron \cite{LNWZ}, HERA \cite{hera}, LEPII \cite{LNPWZ}). Unfortunately, the
range of sparticle and Higgs masses is quite broad, even for the constrained
models we consider. In fact, experimental exploration of all of the well
motivated parameter space of these models would require the large hadron
colliders (LHC/SSC) for the strongly interacting particles (gluino and squarks)
and the next linear collider (NLC) for the weakly interacting particles
(charginos and neutralinos). Of course, the present generation of experiments
has probed and will continue to probe part of this allowed parameter space
(\ie, the lighter end of the spectrum).

Another way of testing the predictions of supergravity models is via indirect
experimental signatures, usually originating from virtual (\eg, one-loop)
processes. In particular, one has the precise electroweak measurements at
LEP (in the form of the $\epsilon_{1,2,3,b}$ parameters
\cite{AB,BFC,ABCI,ABCII}) and the rare radiative decay $b\to s\gamma$
\cite{Bertolini,Barger+Hewett,BG,bsgamma,others}, as observed by CLEOII
Collaboration \cite{Thorndike}. Both these probes have been investigated
independently in the minimal $SU(5)$ and the no-scale flipped $SU(5)$
supergravity models in Ref.~\cite{ewcorr} and \cite{bsgamma}. In this paper we
expand these studies to a larger class of $SU(5)\times U(1)$ supergravity
models and determine the allowed region of parameter space where both
constraints are satisfied {\em simultaneously}.

These one-loop processes have the advantage of side-stepping the strong
kinematical constraints which afflict the direct production channels, although
they require high precision experiments which may not be accountable
exclusively in terms of supersymmetric effects in the models we consider.
Nevertheless, as we point out in this paper, these experimental constraints are
at present the most stringent ones on the class of supergravity models we
consider. It is also important to note that the knowledge acquired through the
indirect tests will help sharpen the predictions for the direct detection
processes, and thus focus the experimental efforts to detect these particles.

This paper is organized as follows. In Sec. 2 we describe briefly the class
of $SU(5)\times U(1)$ supergravity models which we consider, as well as the
four soft-supersymmetry-breaking scenarios of interest. In Sec. 3 we outline
the calculational procedure followed to obtain $\epsilon_1$ and $\bsg$. In
Sec. 4 we present our results and discuss their phenomenological consequences.
Finally in Sec. 5 we summarize our conclusions. The appendix contains the
explicit expressions used to evaluate $\bsg$.

\section{The SU(5) x U(1) Supergravity Models}
\label{models}
The $SU(5)\times U(1)$ supergravity models we consider include the standard
model particles and their superpartners plus two Higgs doublets. These are
supplemented by one additional vector-like quark doublet with mass
$\sim10^{12}\GeV$ and one additional vector-like (charge $-1/3$) quark singlet
with mass $\sim10^6\GeV$ \cite{sism}, such that the unification scale is
delayed until $\sim10^{18}\GeV$ as expected to occur in string models
\cite{Kaplunovsky+Lacaze}.

Besides the two parameters needed to describe the (third-generation) Yukawa
sector of the models ($m_t$ and $\tan\beta$), one must also specify the values
of the soft-supersymmetry-breaking parameters at the unification scale. The
latter can be greatly simplified by assuming universality. That is, all three
gaugino masses are taken to be degenerate ($M_1=M_2=M_3=m_{1/2}$), all fifteen
squark and slepton masses and the two Higgs scalar doublets are separately
assumed to be universal ($M_{(\tilde Q,\tilde U^c,\tilde D^c,\tilde L,\tilde
E^c)_{i=1,2,3}}=m_0$ and $M_{H_{1,2}}=m_0$), and the three trilinear scalar
couplings are taken to be equal ($A_{t,b,\tau}=A$). Of these simplifying
assumptions, only the one relating the (first- and second-generation) scalar
masses is required  experimentally to keep the flavor-changing-neutral-currents
in the $K-\bar K$ system under control \cite{EN}. Nevertheless, these
assumptions are seen to hold in large classes of supergravity models. The
Higgs mixing parameter $\mu$ and its associated bilinear
soft-supersymmetry-breaking mass $B$ do not need to be specified at the
unification scale since they do not feed into the renormalization group
equations for the other parameters. They are determined at the electroweak
scale via the radiative
electroweak symmetry breaking mechanism. In sum then, only five parameters
are needed to describe the supergravity models we consider. In contrast,
twenty-five parameters would be needed in the MSSM.

The minimization of the electroweak Higgs potential is performed at the
one-loop level, and yields the values of $|\mu|$ and $B$ as well as the
one-loop corrected Higgs boson masses \cite{aspects}. The five-dimensional
parameter space is restricted by several consistency conditions and
experimental constraints on the unobserved sparticle masses (most importantly
the chargino mass and the lightest Higgs boson mass \cite{LNPWZh}). Moreover,
the cosmological constraint of a not-too-young universe (usually incorrectly
refered to as ``not overclosing the universe") is also imposed \cite{KLNPYdm}.

Building on the above basic structure one is naturally led to the $SU(5)\times
U(1)$ (or flipped $SU(5)$) gauge group since this model has a strong motivation
in the context of string theory \cite{EriceDec92}. Moreover, the additional
particles fit snugly into complete \r{10},\rb{10} $SU(5)\times U(1)$
representations. In this model proton decay and cosmological constraints are
satisfied automatically \cite{LNZI,LNZII}. Concerning the
soft-supersymmetry-breaking parameter space, one assumes two possible
string-inspired supersymmetry breaking scenarios:
\begin{description}
\item (i) the $SU(N,1)$ no-scale supergravity model \cite{nsII,LN} which
implies $m_0=A=0$, with a more constrained case called the strict no-scale
scenario where $B(M_U)=0$ is also assumed; and
\item (ii) the dilaton supersymmetry breaking scenario \cite{KL} where
$m_0=\coeff{1}{\sqrt{3}}m_{1/2}$ and $A=-m_{1/2}$, including a special case
where $B(M_U)=2m_0$.
\end{description}
The allowed parameter spaces in these two cases have been determined in
Refs.~\cite{LNZI} and \cite{LNZII} respectively. It is found that there are no
important constraints on $\tan\beta$ ($\tan\beta\lsim32,46$ respectively) and
the sparticle masses can be as light as their present experimental lower
bounds. More importantly, the parameter spaces are three-dimensional
($m_t,\tan\beta,m_{1/2}\propto m_{\tilde g}$) and therefore the spectrum scales
with $m_{\tilde g}$ \cite{EriceDec92}. (In what follows we consider only three
representative values of $m_t=130,150,170\GeV$.) In Table \ref{Table1} we
give a comparative listing of the most important properties of these models.

It is interesting to note that the special cases of these two supersymmetry
breaking scenarios are quite predictive, since they allow to determine
$\tan\beta$ in terms of $m_t$ and $m_{\tilde g}$. In the strict no-scale case
one finds that the value of $m_t$ determines the sign of $\mu$
($\mu>0:\,m_t\lsim135\GeV$, $\mu<0:\,m_t\gsim140\GeV$) and whether the lightest
Higgs boson mass is above or below $100\GeV$. In the special dilaton scenario,
$\tan\beta\approx1.4-1.6$ and $m_t\lsim155\GeV$, $61\GeV\lsim m_h\lsim91\GeV$
follow. Thus, continuing Tevatron top-quark searches and LEPI,II Higgs searches
could probe these restricted scenarios completely.

\section{One-loop Constraints ($\epsilon_1$ and $b\to s\gamma$)}
The one-loop corrections to the $W^\pm$ and $Z$ boson self-energies
(\ie, the ``oblique" corrections) can be parametrized in terms of three
variables $\epsilon_{1,2,3}$ \cite{AB} which are constrained experimentally by
the precise LEP measurements of the $Z$ leptonic width ($\Gamma_l$), and the
leptonic forward-backward asymmetries at the $Z$-pole ($A^l_{FB}$), as well
as the $M_W/M_Z$ ratio. A fourth observable is the $Z\to b\bar b$ width which
is described by the $\epsilon_b$ parameter \cite{ABCI}. Of these four
variables, at present $\epsilon_1$ provides the strongest constraint in
supersymmetric models at the 90\%CL \cite{ewcorr,ABCII}. However, the
$\epsilon_b$ constraint is competitive with (although at present somewhat
weaker than) the $\epsilon_1$ constraint \cite{ABCII,LNPZII}, and in fact
may impose interesting constraints on supersymmetric models as the precision
of the data increases.

The expression for $\epsilon_1$ is obtained from the following definition
\cite{BFC}
\beq
\epsilon_1=e_1-e_5-{\delta G_{V,B}\over G}-4\delta g_A,\label{eps1}
\eeq
where $e_{1,5}$ are the following combinations of vacuum polarization
amplitudes
\begin{eqnarray}
e_1&=&{\alpha\over 4\pi \sin^2\theta_W M^2_W}[\Pi^{33}_T(0)-\Pi^{11}_T(0)],
\label{e1}\\
e_5&=& M_Z^2F^\prime_{ZZ}(M_Z^2),\label{e5}
\end{eqnarray}
and the $q^2\not=0$ contributions $F_{ij}(q^2)$ are defined by
\beq
\Pi^{ij}_T(q^2)=\Pi^{ij}_T(0)+q^2F_{ij}(q^2).
\eeq
The $\delta g_A$ in Eqn.~(\ref{eps1}) is the contribution to the axial-vector
form factor at $q^2=M^2_Z$ in the $Z\to l^+l^-$ vertex from proper vertex
diagrams and fermion self-energies, and $\delta G_{V,B}$ comes from the
one-loop box, vertex and fermion self-energy corrections to the $\mu$-decay
amplitude at zero external momentum. These non-oblique SM corrections are
non-negligible, and must be included in order to obtain an accurate SM
prediction.

As is well known, the SM contribution to $\epsilon_1$ depends quadratically
on $m_t$ but only logarithmically on the SM Higgs boson mass ($m_H$). In this
fashion upper bounds on $m_t$ can be obtained which have a non-negligible $m_H$
dependence: up to $20\GeV$ stronger when going from a heavy ($\approx1\TeV$)
to a light ($\approx100\GeV$) Higgs boson. It is also known (in the MSSM) that
the largest supersymmetric contributions to $\epsilon_1$ are expected to
arise from the $\tilde t$-$\tilde b$ sector, and in the limiting case of a very
light stop, the contribution is comparable to that of the $t$-$b$ sector. The
remaining squark, slepton, chargino, neutralino, and Higgs sectors all
typically contribute considerably less. For increasing sparticle masses, the
heavy sector of the theory decouples, and only SM effects  with a {\it light}
Higgs boson survive. (This entails stricter upper bounds on $m_t$ than in the
SM, since there the Higgs boson does not need to be light.) However, for a
light chargino ($m_{\chi^\pm_1}\to{1\over2}M_Z$), a $Z$-wavefunction
renormalization threshold effect can introduce a substantial $q^2$-dependence
in the calculation, \ie, the presence of $e_5$ in Eq.~(\ref{eps1}) \cite{BFC}.
The complete vacuum polarization contributions from the Higgs sector, the
supersymmetric chargino-neutralino and sfermion sectors, and also the
corresponding contributions in the SM have been included in our calculations
\cite{ewcorr}.

The rare radiative flavor-changing-neutral-current (FCNC) $b\to s\gamma$ decay
has been observed by the CLEOII Collaboration in the following 95\% CL allowed
range\cite{Thorndike}
\beq
\bsg=(0.6-5.4)\times10^{-4}.\label{bsg}
\eeq
In Ref.~\cite{bsgamma} we have given the predictions for the branching ratio
in the minimal $SU(5)$ supergravity model
($\bsg_{minimal}=(2.3-3.6)\times10^{-4}$) and in the no-scale flipped $SU(5)$
supergravity model. However, in that paper the experimental {\em lower} bound
on $\bsg$ was not available. Since a large suppression of $\bsg$ (much below
the SM value) can occur in the flipped $SU(5)$ models, such a bound can be
quite restrictive. Below we give the predictions for $\bsg$ in the two variants
of the flipped $SU(5)$ model (and their special subcases) described in
Sec.~\ref{models}. The expressions used to compute the branching ratio $\bsg$
are given in the Appendix for completeness.

\section{Discussion}
The results of our computations for $\bsg$ and $\epsilon_1$ are shown in
Figures 1,3,4,6 in the various models under consideration, for
$m_t=130,150,170\GeV$. (Smaller values of $m_t$ are not ruled out
experimentally, although they appear ever more unlikely.) The LEP value for
$\epsilon_1$ which we use in our analysis is
$\epsilon_1=(-0.9\pm3.7)\times10^{-3}$ \cite{BLec}, which implies
$\epsilon_1<0.00517\,(0.00632)$ at the 90(95)\%CL. The experimentally allowed
interval for $\bsg$ is given in Eq.~(\ref{bsg}), although in what follows we
will also consider a less conservative estimate of the lower bound, namely
$\bsg>10^{-5}$. We now discuss the results and ensuing constraints on each
model in turn.
\subsection{No-scale flipped SU(5)}
At the 90(95)\%CL, for $m_t\lsim150\,(165)\GeV$ there are no restrictions on
the model parameters from the $\epsilon_1$ constraint (see Fig. 1). For
$m_t=170\GeV$ (see Fig. 1c) the $\epsilon_1$ constraint {\em alone} implies a
strict upper bound on the chargino mass:\footnote{An upper bound on the
chargino mass implies upper bounds on {\em all} sparticle and Higgs masses,
since they are all proportional to $m_{\tilde g}$, see Table~\ref{Table1}.} (i)
for $\mu>0$ there are no allowed points at 90\%CL, while
$m_{\chi^\pm_1}\lsim70\GeV$ is required at 95\%CL; (ii) for $\mu<0$ one obtains
$m_{\chi^\pm_1}\lsim58\,(70)\GeV$ at 90(95)\%CL. Interestingly enough, for this
range of chargino masses the $\bsg$ constraint is also restrictive. {\em
Combining} the $\epsilon_1$ and $\bsg$ constraints we obtain:
(i) for $\mu>0$, $m_{\chi^\pm_1}\lsim67\GeV$ at 95\%CL and
$\tan\beta\approx8-10$; (ii) for $\mu<0$, $m_{\chi^\pm_1}\lsim54\,(67)\GeV$ at
90(95)\%CL and $\tan\beta\lsim8$.
No significant improvement is obtained by required $\bsg>10^{-5}$. Analogously,
upper bounds on $m_{\chi^\pm_1}$ up to $\approx100\GeV$ are obtained for values
of $m_t$ in the range $150-170\GeV$ \cite{ewcorr}. Larger values of $m_t$
($m_t\lsim190\GeV$ is required \cite{LNZI}) could only be made consistent with
LEP data if the chargino mass is very near its present experimental lower
bound.

For $m_t=130,150\GeV$, the $\epsilon_1$ constraint is ineffective. However,
for $\mu>0$ the $\bsg$ constraint is quite restrictive, as shown in Figs.
1a,1b. The various (dotted) curves correspond to different values of
$\tan\beta$. For large values of $m_{\chi^\pm_1}$, these curves start off at
values of $\bsg$ which decrease with increasing $\tan\beta$, \ie, the largest
value corresponds to $\tan\beta=2$, and then $\tan\beta$ increases in steps of
two. As the chargino mass decreases, these curves reach a minimum (\ie, zero)
value and then increase again (except for $\tan\beta=2$), and even exceed
the upper bound on $\bsg$ for large enough $\tan\beta$. To show better the
excluded area, in Fig. 2 we have plotted those points in parameter space which
survive the $\bsg$ constraint, in the $(m_{\chi^\pm_1},\tan\beta)$ plane.
The swath along the diagonal is excluded because $\bsg$ is too small. In fact,
if we demand $\bsg>10^{-5}$, the points denoted by crosses would be excluded
as well. The area to the left of the left group of points is excluded because
$\bsg$ is too large. Note that no matter what the actual value of $\bsg$ ends
up being, there will always be some allowed set of points, namely, a subset
of {\em both} sets of presently allowed points. Another consequence of the
$\bsg$ constraint is an upper bound on $\tan\beta$: for $m_t=130\,(150)\GeV$,
$\tan\beta\lsim26\,(20)$ compared to the upper bound of
$\tan\beta\lsim30\,(26)$ which existed prior to the application of the $\bsg$
constraint. Finally, note that $\tan\beta\gsim20$ implies
$m_{\chi^\pm_1}\gsim100\GeV$.

\subsection{Strict no-scale flipped SU(5)}
In this variant of the model $\tan\beta$ is determined for given
$m_t$ and $m_{\tilde g}$ values and is such that $m_t=130\GeV$ is only allowed
for $\mu>0$, whereas $m_t=150,170\GeV$ are only allowed for $\mu<0$ (see
Table~\ref{Table1} and Ref.~\cite{LNZI}). As above, the $\epsilon_1$ constraint
is only effective for $m_t\gsim150\,(165)\GeV$ at the 90(95)\%CL. For
$m_t=170\GeV$ there is an upper bound $m_{\chi^\pm_1}\lsim54\,(62)\GeV$ at
90(95)\%CL, and there is no further constraint from $\bsg$ (see Fig. 3). In
fact, $\bsg$ is only constraining for $m_t=130\GeV$. Note that in this case
there are two possible solutions for $\tan\beta$ which are most clearly seen in
the $\bsg$ plot (Fig. 3). The lower $\tan\beta$ solution (which asymptotes to
the larger value of $\bsg$) excludes chargino masses in the interval
$95\GeV\lsim m_{\chi^\pm_1}\lsim175\GeV$, whereas the larger $\tan\beta$
solution excludes $m_{\chi^\pm_1}\gsim150\GeV$.

\subsection{Dilaton flipped SU(5)}
The discussion in this model parallels closely that for the no-scale model
given above, with some relevant differences. For $m_t=170\GeV$, the
$\epsilon_1$ constraint alone implies (see Fig. 4c): (i) for $\mu>0$,
$m_{\chi^\pm_1}\lsim53\,(60)\GeV$ at the 90(95)\%CL and $\tan\beta\lsim10$, and
(ii) for $\mu<0$, $m_{\chi^\pm_1}\lsim56\,(70)\GeV$ at the 90(95)\%CL and
$\tan\beta\lsim24$. The further imposition of the $\bsg$ constraint is not
significant.

For $m_t=130,150\GeV$, the $\bsg$ constraint is quite restrictive, and this
time for {\em both} signs of $\mu$ (see Figs. 4a,4b), although it is more
important for $\mu>0$. The allowed regions of parameter space in the
$(m_{\chi^\pm_1},\tan\beta)$ space are shown in Fig. 5. The only qualitative
difference with the results for the no-scale case is that for $m_t=150\GeV$ and
$\mu<0$, $\bsg$ can only be too small, and this excludes the area to the left
of the set of allowed points. Finally, $\bsg$ does not impose any further
constraints on $\tan\beta$.

\subsection{Special dilaton flipped SU(5)}
In this case, only $\mu<0$ is allowed  and $m_t\lsim155\GeV$ is required. It
also follows that $\tan\beta<2$ (see Table~\ref{Table1} and Ref.~\cite{LNZII}).
At the moment $\bsg$ does not impose any constraints on the parameter space
(see Fig. 6). The same is true for $\epsilon_1$, except for $m_t=155\GeV$ which
at the 90\%CL cannot be made acceptable since light chargino masses are not
allowed (see Fig. 6); there are no constraints at the 95\%CL.

\section{Conclusions}
We have considered a class of well motivated string-inspired supergravity
models based on the gauge group $SU(5)\times U(1)$. The various constraints
on the models allow one to predict all new low-energy phenomena in terms of
just three parameters ($m_t,\tan\beta,m_{\tilde g}$) and as such these models
are highly predictive. We have shown that one-loop processes which do not
create real sparticles can nonetheless be used to constrain these models
in interesting ways. The LEP $\epsilon_1$ constraint and the CLEOII
$\bsg$ allowed range are perhaps surprisingly restrictive in the models which
we have considered. The $\epsilon_1$ constraint requires the presence of light
charginos if the top-quark mass exceeds the moderate value of $\approx155\GeV$.
In fact, $m_{\chi^\pm_1}\lsim100\GeV$ is the weakest possible constraint in
this case. Since all sparticle masses are proportional to $m_{\tilde g}$,
$\tan\beta$-dependent upper bounds on all of them also follow. The weakest
possible upper bounds (\ie, $\tan\beta$-independent) which follow for
$m_t\gsim155\,(165)\GeV$ at the 90(95)\%CL are:
\begin{eqnarray}
&m_{\chi^0_1}\lsim55\GeV,\qquad\qquad&m_{\chi^\pm_1,\chi^0_2}\lsim100\GeV,\\
&m_{\tilde g}\lsim360\GeV,\qquad\qquad&m_{\tilde q}\lsim350\,(365)\GeV,\\
&\!\!\!m_{\tilde e_R}\lsim80\,(125)\GeV,
\quad&m_{\tilde e_L}\lsim120\,(155)\GeV,\quad m_{\tilde\nu}\lsim100\,(140)\GeV
\end{eqnarray}
in the no-scale (dilaton) flipped $SU(5)$ supergravity model.

The $\bsg$ constraint can probe the models irrespective of the mass scales
involved because of a large suppression of the amplitude for a wide range of
chargino masses ($50\GeV\lsim m_{\chi^\pm_1}\lsim280\GeV$ depending on the
value of $\tan\beta$). It should be noted that most of the qualitative aspects
of our discussions should apply quite generally to the class of supergravity
models with radiative electroweak symmetry breaking, where the parameters $m_0$
and $A$ are not necessarily related to $m_{1/2}$ in the ways discussed here.

We conclude that future refinements of the allowed $\bsg$ range are likely to
result in decisive tests of this class of models. Moreover, if the top-quark
mass is too heavy to be easily detectable at the Tevatron, then the
$\epsilon_1$ constraint would require light charginos. These could nevertheless
escape detection at the Tevatron, since the characteristic trilepton events are
generally suppressed for light charginos \cite{LNWZ}. However, the outlook for
chargino and selectron detection at LEPII \cite{LNPWZ} and at HERA \cite{hera}
will remain quite favorable in this class of models.

\section*{Acknowledgements}
This work has been supported in part by DOE grant DE-FG05-91-ER-40633. The work
of J.L. has been supported by an SSC Fellowship.

\appendix
\section{Expression for $\bsg$}
The expression used for $\bsg$ is given by \cite{BG}
\beq
{B(b\to s\gamma)\over B(b\to ce\bar\nu)}={6\alpha\over\pi}
{\left[\eta^{16/23}A_\gamma
+\coeff{8}{3}(\eta^{14/23}-\eta^{16/23})A_g+C\right]^2\over
I(m_c/m_b)\left[1-\coeff{2}{3\pi}\alpha_s(m_b)f(m_c/m_b)\right]},
\eeq
where $\eta=\alpha_s(M_Z)/\alpha_s(m_b)$, $I$ is the phase-space factor
$I(x)=1-8x^2+8x^6-x^8-24x^4\ln x$, and $f(m_c/m_b)=2.41$ the QCD
correction factor for the semileptonic decay. In our computations we have used:
$\alpha_s(M_Z)=0.118$, $ B(b\to ce\bar\nu)=10.7\%$, $m_b=4.8\GeV$, and
$m_c/m_b=0.3$. The $A_\gamma,A_g$ are the
coefficients of the effective $bs\gamma$ and $bsg$ penguin operators
evaluated at the scale $M_Z$. Their simplified expressions are given below,
in the justifiable limit of negligible gluino and neutralino contributions
\cite{Bertolini} and degenerate squarks, except for the $\tilde t_{1,2}$ which
are significantly split by $m_t$.

The contributions to $A_{\gamma ,g}$ from the $W-t$ loop, the $H^\pm-t$ loop,
and the $\chi^\pm_i-\tilde t_k$ loop are given by
\begin{eqnarray}
W:&& A_{\gamma ,g} =\frac{3}{2}\frac{m_t^2}{m_W^2}f^{(1)}_{\gamma ,g}
\left( \frac{m_t^2}{m_W^2}\right) \\
{H^\pm}:&& A_{\gamma ,g} =\frac{1}{2}\frac{m_t^2}{m_{H^\pm}^2}\left[
\frac{1}{\tan^2 \beta}
f^{(1)}_{\gamma ,g} \left( \frac{m_t^2}{m_{H^\pm}^2}\right) +
f^{(2)}_{\gamma ,g} \left( \frac{m_t^2}{m_{H^\pm}^2}\right) \right] \\
\chi^\pm_i :&& A_{\gamma ,g} =\sum_{j=1}^{2} \left\{
\frac{m_W^2}{m_{\chi^\pm_j}^2}\left[ |V_{j1}|^2
f^{(1)}_{\gamma ,g} \left( \frac{m_{\tilde Q}^2}{m_{\chi^\pm_j}^2}\right)
\right. \right. \nonumber \\
&&-\left. \sum_{k=1}^2 \left| V_{j1}T_{k1}-V_{j2}T_{k2}\frac{m_t}{\sqrt{2}
m_W \sin \beta} \right|^2
f^{(1)}_{\gamma ,g} \left( \frac{m_{\tilde t_k}^2}
{m_{\chi^\pm_j}^2}\right) \right] \nonumber \\
&&-\frac{U_{j2}}{\sqrt{2} \cos \beta}
\frac{m_W}{m_{\chi^\pm_j}}\left[ V_{j1}
f^{(3)}_{\gamma ,g} \left( \frac{m_{\tilde Q}^2}{m_{\chi^\pm_j}^2}\right)
\right. \nonumber \\
&&-\left. \left.
\sum_{k=1}^2 \left( V_{j1}T_{k1}-V_{j2}T_{k2}\frac{m_t}{\sqrt{2}
m_W \sin \beta} \right) T_{k1}
f^{(3)}_{\gamma ,g} \left( \frac{m_{\tilde t_k}^2}
{m_{\chi^\pm_j}^2}\right) \right]  \right\},
\end{eqnarray}
with
\begin{eqnarray}
f^{(1)}_\gamma (x) &=& \frac{(7-5x-8x^2)}{36(x-1)^3}+
\frac{x(3x-2)}{6(x-1)^4}\log x \\
f^{(2)}_\gamma (x) &=& \frac{(3-5x)}{6(x-1)^2}+
\frac{(3x-2)}{3(x-1)^3}\log x \\
f^{(3)}_\gamma (x) &=& (1-x) f^{(1)}_\gamma (x) -
\frac{x}{2}f^{(2)}_\gamma (x) -\frac{23}{36} \\
f^{(1)}_g (x) &=& \frac{(2+5x-x^2)}{12(x-1)^3}-
\frac{x}{2(x-1)^4}\log x \\
f^{(2)}_g (x) &=& \frac{(3-x)}{2(x-1)^2}-
\frac{1}{(x-1)^3}\log x \\
f^{(3)}_g (x) &=& (1-x) f^{(1)}_g (x) -
\frac{x}{2}f^{(2)}_g (x) -\frac{1}{3}.
\end{eqnarray}

In these formulas $U_{ij},V_{ij}$ are the elements of the matrices which
diagonalize the chargino mass matrix, $m_{\chi^\pm_i}$ are the chargino masses,
$m_{\tilde t_k}$ are the stop mass eigenvalues, $T_{ij}$ are the elements
of the matrix which diagonalizes the $2\times2$ stop mass matrix.

\begin{table}[p]
\hrule
\caption{Major features of the class of $SU(5)\times U(1)$ supergravity models,
and a comparison of the supersymmetry breaking scenarios considered.
(All masses in GeV).}
\label{Table1}
\begin{center}
\begin{tabular}{|l|}\hline
\hfil$\bf SU(5)\times U(1)$\hfil\\ \hline
$\bullet$ Easily string-derivable, several known examples\\
$\bullet$ Symmetry breaking to Standard Model due to vevs of \r{10},\rb{10}\\
\quad and tied to onset of supersymmetry breaking\\
$\bullet$ Natural doublet-triplet splitting mechanism\\
$\bullet$ Proton decay: $d=5$ operators very small\\
$\bullet$ Baryon asymmetry through lepton number asymmetry\\
\quad (induced by the decay of flipped neutrinos) as recycled by\\
\quad non-perturbative electroweak interactions\\
\hline
\end{tabular}
\end{center}
\begin{center}
\begin{tabular}{|l|l|}\hline
\hfil No-Scale\hfil&\hfil Dilaton\hfil\\ \hline
$\bullet$ Parameters 3: $m_{1/2},\tan\beta,m_t$&
$\bullet$ Parameters 3: $m_{1/2},\tan\beta,m_t$\\
$\bullet$ Universal soft-supersymmetry-&$\bullet$ Universal
soft-supersymmetry-\\
\quad breaking automatic&\quad breaking automatic\\
$\bullet$ $m_0=0$, $A=0$&$\bullet$ $m_0=\coeff{1}{\sqrt{3}}m_{1/2}$,
$A=-m_{1/2}$\\
$\bullet$ Dark matter: $\Omega_\chi h^2_0<0.25$
&$\bullet$ Dark matter: $\Omega_\chi h^2_0<0.90$\\
$\bullet$ $m_{1/2}<475\GeV$, $\tan\beta<32$
&$\bullet$ $m_{1/2}<465\GeV$, $\tan\beta<46$\\
$\bullet$ $m_{\tilde g}>245\GeV$, $m_{\tilde q}>240\GeV$
&$\bullet$ $m_{\tilde g}>195\GeV$, $m_{\tilde q}>195\GeV$\\
$\bullet$ $m_{\tilde q}\approx0.97m_{\tilde g}$
&$\bullet$ $m_{\tilde q}\approx1.01m_{\tilde g}$\\
$\bullet$ $m_{\tilde t_1}>155\GeV$
&$\bullet$ $m_{\tilde t_1}>90\GeV$\\
$\bullet$ $m_{\tilde e_R}\approx0.18m_{\tilde g}$,
$m_{\tilde e_L}\approx0.30m_{\tilde g}$
&$\bullet$ $m_{\tilde e_R}\approx0.33m_{\tilde g}$,
$m_{\tilde e_L}\approx0.41m_{\tilde g}$\\
\quad $m_{\tilde e_R}/m_{\tilde e_L}\approx0.61$
&\quad $m_{\tilde e_R}/m_{\tilde e_L}\approx0.81$\\
$\bullet$ $60\GeV<m_h<125\GeV$&$\bullet$ $60\GeV<m_h<125\GeV$\\
$\bullet$ $2m_{\chi^0_1}\approx m_{\chi^0_2}\approx m_{\chi^\pm_1}\approx
0.28 m_{\tilde g}\lsim290$
&$\bullet$ $2m_{\chi^0_1}\approx m_{\chi^0_2}\approx m_{\chi^\pm_1}\approx
0.28 m_{\tilde g}\lsim290$\\
$\bullet$ $m_{\chi^0_3}\sim m_{\chi^0_4}\sim m_{\chi^\pm_2}\sim\vert\mu\vert$
&$\bullet$ $m_{\chi^0_3}\sim m_{\chi^0_4}\sim m_{\chi^\pm_2}\sim\vert\mu\vert$
\\
$\bullet$ Spectrum easily accessible soon
&$\bullet$ Spectrum accessible soon\\ \hline
$\bullet$ Strict no-scale: $B(M_U)=0$
&$\bullet$ Special dilaton: $B(M_U)=2m_0$\\
\quad $\tan\beta=\tan\beta(m_t,m_{\tilde g})$
&\quad $\tan\beta=\tan\beta(m_t,m_{\tilde g})$\\
\quad $m_t\lsim135\GeV\Rightarrow\mu>0,m_h\lsim100\GeV$
&\quad $\tan\beta\approx1.4-1.6$, $m_t<155\GeV$\\
\quad $m_t\gsim140\GeV\Rightarrow\mu<0,m_h\gsim100\GeV$
&\quad $m_h\approx61-91\GeV$\\
\hline
\end{tabular}
\end{center}
\hrule
\end{table}
\newpage

\noindent{\large\bf Figure Captions}
\begin{description}
\item Figure 1: The values of $\bsg$ (top row) and the $\epsilon_1$ parameter
(bottom row) versus the chargino mass in no-scale flipped $SU(5)$ supergravity
model for (a) $m_t=130\GeV$, (b) $m_t=150\GeV$, and (c) $m_t=170\GeV$. The 95\%
CL CLEOII limit on $\bsg$ and the 90\% CL LEP upper limit on $\epsilon_1$ are
indicated.
\item Figure 2: The region in $(m_{\chi^\pm_1},\tan\beta)$ space which is
allowed by the CLEOII limit on $\bsg$ in the no-scale flipped $SU(5)$
supergravity model, for $\mu>0$ and $m_t=130,150\GeV$. The value of $\bsg$
is too large to the left of the left group of points, and too small in between
the two groups of points. (No such constraints exist for $\mu<0$.) The points
denoted by crosses would become excluded if the lower bound on $\bsg$ is
strenghthened to $\bsg>10^{-5}$.
\item Figure 3: The values of $\bsg$ (top row) and the $\epsilon_1$ parameter
(bottom row) versus the chargino mass in the strict no-scale flipped $SU(5)$
supergravity model, for $m_t=130\GeV$ ($\mu>0$) and $m_t=150,170\GeV$
($\mu<0$). The 95\% CL CLEOII limit on $\bsg$ and the 90\% CL LEP upper limit
on $\epsilon_1$ are indicated. For $m_t=130\GeV$ ($\mu>0$) two $\tan\beta$
solutions exist which are clearly visible in the $\bsg$ plot.
\item Figure 4: The values of $\bsg$ (top row) and the $\epsilon_1$ parameter
(bottom row) versus the chargino mass in the dilaton flipped $SU(5)$
supergravity model, for (a) $m_t=130\GeV$, (b) $m_t=150\GeV$, and (c)
$m_t=170\GeV$. The 95\% CL CLEOII limit on $\bsg$ and the 90\% CL LEP upper
limit on $\epsilon_1$ are indicated.
\item Figure 5: The region in $(m_{\chi^\pm_1},\tan\beta)$ space which is
allowed by the CLEOII limit on $\bsg$ in the dilaton flipped $SU(5)$
supergravity model, for $m_t=130,150\GeV$. The value of $\bsg$ is too large to
the left of the left group of points, and too small in between the two groups
of points (except for $m_t=150\GeV$ ($\mu<0$), where it is too small to the
left of the single group of points). The points denoted by crosses would become
excluded if the lower bound on $\bsg$ is strenghthened to $\bsg>10^{-5}$.
\item Figure 6: The values of $\bsg$ (top) and the $\epsilon_1$ parameter
(bottom) versus the chargino mass in the special dilaton $SU(5)$ supergravity
model (only allowed for $\mu<0$) for $m_t=130,150,155\GeV$. (In this model
$m_t\lsim155\GeV$ is required.) The 95\% CL CLEOII limit on $\bsg$ and the 90\%
CL LEP upper limit on $\epsilon_1$ are indicated.
\end{description}

\end{document}